\documentclass[conference]{IEEEtran}

\IEEEoverridecommandlockouts
\usepackage{cite}
\usepackage{amsmath,amssymb,amsfonts, amsthm}
\usepackage{algorithm}
\usepackage{multirow}
\usepackage{algpseudocode}
\usepackage{graphicx}
\usepackage{graphicx}
\usepackage{textcomp}
\usepackage{xcolor}
\usepackage{soul}
\def\BibTeX{{\rm B\kern-.05em{\sc i\kern-.025em b}\kern-.08em
    T\kern-.1667em\lower.7ex\hbox{E}\kern-.125emX}}

\newtheorem{lemma}{Lemma}
\newtheorem{proposition}{Proposition}
\begin{document}

\title{Survival of the Optimized: An Evolutionary Approach to $T$-depth Reduction\\

}

\author{\IEEEauthorblockN{Archisman Ghosh}
\IEEEauthorblockA{\textit{CSE Department} \\
\textit{Pennsylvania State University}\\
State College, PA, USA\\
apg6127@psu.edu}

\and
\IEEEauthorblockN{Avimita Chatterjee}
\IEEEauthorblockA{\textit{CSE Department} \\
\textit{Pennsylvania State University}\\
State College, PA, USA \\
amc8313@psu.edu}
\and
\IEEEauthorblockN{Swaroop Ghosh}
\IEEEauthorblockA{\textit{School of EECS} \\
\textit{Pennsylvania State University}\\
State College, PA, USA \\
szg212@psu.edu}}
% \and
% \and
% \and

\maketitle

\begin{abstract}
Quantum Error Correction (QEC) is the cornerstone of practical Fault-Tolerant Quantum Computing (FTQC), but incurs enormous resource overheads. Circuits must decompose into Clifford+$T$ gates, and the non-transversal $T$ gates demand costly magic-state distillation. As circuit complexity grows, sequential $T$-gate layers (``$T$-depth”) increase, amplifying the spatiotemporal overhead of QEC. Optimizing $T$-depth is NP-hard, and existing greedy or brute-force strategies are either inefficient or computationally prohibitive. We frame $T$-depth reduction as a search optimization problem and present a Genetic Algorithm (GA) framework that approximates optimal layer-merge patterns across the non-convex search space. We introduce a mathematical formulation of the circuit expansion for systematic layer reordering and a greedy initial merge-pair selection, accelerating the convergence and enhancing the solution quality. In our benchmark with $\sim$90–100 qubits, our method reduces $T$-depth by 79.23\% and overall $T$-count by 41.86\%. Compared to the standard reversible circuit benchmarks, we achieve a $\sim 2.58 \times$ average improvement in $T$-depth over the state-of-the-art methods, demonstrating its viability for near-term FTQC.

\end{abstract}


\section{Introduction}

In the Noisy Intermediate-Scale Quantum (NISQ) era \cite{preskill1998reliable}, $\sim100$-qubit devices solve classically intractable tasks but remain highly error-prone due to limited coherence times and environmental noise. Practical quantum error correction is thus essential for near-term fault-tolerant quantum computing (FTQC). The surface code—with local qubit interactions and a high error threshold of $0.7\%$—is the most feasible candidate; when paired with a magic state factory, it enables universal Clifford+$T$ computation \cite{kliuchnikov2012fast,brown2017poking}. In this model, Clifford gates can be absorbed into measurement bases (e.g., a Hadamard before measurement converts an $X$ measurement into $Z$) \cite{gottesman1998heisenberg}, simplifying their implementation. In contrast, $T$ gates require magic state injection: the state \( |m\rangle = |0\rangle + e^{i\pi/4}|1\rangle \) must be injected into the code and distilled to achieve the high fidelity needed for fault tolerance \cite{littinski}.

The fidelity of initially prepared magic states critically determines the resource cost of distillation \cite{haah2018codes,bravyi2005universal}. In the 15-to-1 protocol, an injection error rate \(p\) is suppressed to \(\sim p^3\) after one round, so reducing \(p\) from \(10^{-3}\) to \(10^{-4}\) yields a three-order-of-magnitude improvement in distilled fidelity. This nonlinear amplification means that the distinction between \(p\sim10^{-3}\) and \(p\sim10^{-4}\) can decide whether a single round suffices or if multiple rounds are required to meet near-term FTQC thresholds \cite{litinski2019magic}. Each additional round multiplies resource consumption, exacerbating hardware demands. Moreover, the 15-to-1 protocol employs 11 tiles—each occupying \(2d^2-1\) physical qubits—for a total of 241 qubits \cite{littinski}. Consequently, pruning 100 \(T\)-gates in a 10-qubit circuit can demand on the order of \(10^7\) qubits, creating a primary bottleneck for near-term fault-tolerant quantum computers.

\begin{figure*}
    
    \centering
    \includegraphics[width=\linewidth]{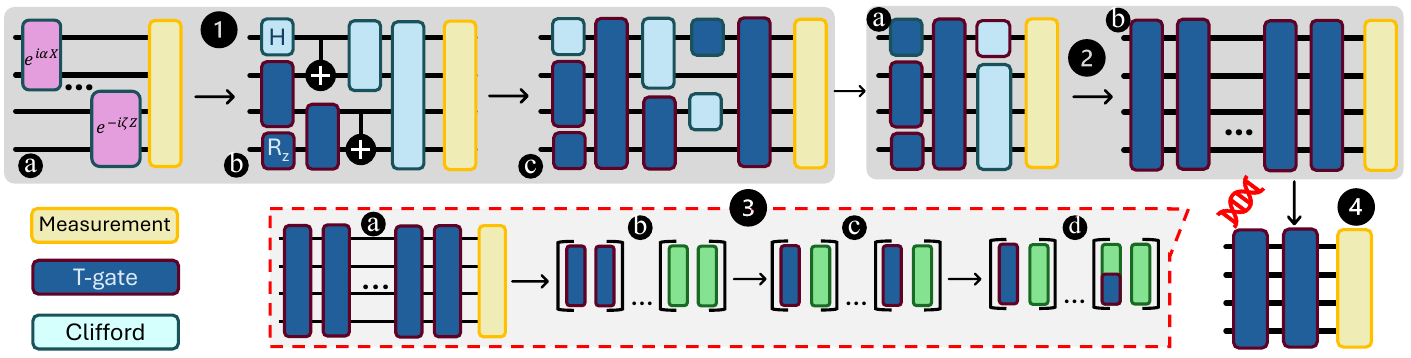}
    %\vspace{-10pt}
    \caption{The diagram describes the overall flow of Fault-Tolerant Quantum Computing (FTQC). (1) is the initial transformation of the circuit for the application of a QEC protocol. 1(a) is the user-designed quantum circuit; 1(b) represents the transpiled circuit in the native gate set of the execution hardware; and 1(c) is the Clifford+$T$ formalism of the circuit for a QEC protocol, like surface code. 2(a) represents the transformation of the Clifford gates to a sequence of Pauli rotations and are commuted past the $T$-gates through certain commutation rules \cite{littinski}; and 2(b) shows the un-optimized circuit with layers of $T$-gates which serve as the input to our optimization framework. (3) illustrates the optimization of the $T$-depth which is the primary contribution of the paper. 3(a) describes the initial population for the GA; 3(b) is the merge pairs chosen through a greedy procedure detailed in Section IV; 3(c) and 3(d) represent the selection, crossover, and mutation stages which ultimately result in (4) which is the optimized circuit with a reduced $T$-depth.}
    \label{fig:flow}
    % \vspace{-10pt}
\end{figure*}
Therefore, reducing $T$-depth by packing commuting $\pi/8$ rotations into as few sequential layers as possible, provides a solution to the bottleneck by directly reducing the number of fresh magic states, and thus the qubit overhead, needed for magic-state distillation.
% The immense qubit overhead for implementing a magic state distillation protocol (15-to-1) for the pruning of $T$-gate layers in the circuit is a primary bottleneck in the implementation of FTQC. Magic states to be produced using the 15-to-1 protocol requires 11 tiles \cite{littinski} places an overhead of 241 qubits ($2d^2-1$ qubits for $d$ tiles). Therefore, for a small circuit of 10 qubits and 100 $T$ gates, the qubit overhead for pruning the non-Clifford gates can be as high as the order of $10^7$, which is extremely high for near-term fault-tolerant quantum computers.

% Based on the Clifford+$T$ formalism, a $n$ qubit circuit can be represented as a number of consecutive $n$ final Pauli product measurements and consecutive $\pi/8$ rotation gates, which can be grouped into layers that commute with each other. This refers to the $T$-depth of a circuit. Based on the rules of commutativity in rotation gates, the $T$-depth of a circuit can be reduced by grouping such layers. Since, a singular magic state is required to prune a $T$-gate layer, combining the layers and ultimately reducing the $T$-depth of a circuit leads to the usage of a lesser number of magic states to absorb the $T$ gates. This reduces the qubit overhead required in the costly magic state distillation procedure.
In this paper, we propose a framework to optimize Clifford+$T$ circuits by merging commuting $T$-gate layers to reduce magic-state distillation overhead.  We formulate layer merging as a search-optimization problem and employ a genetic algorithm (GA) initialized with the Clifford-pruned circuit (Fig.~\ref{fig:flow}.3(a)).  Layer pairs are selected by $T$-gate density (Fig.~\ref{fig:flow}.3(b)), and GA operators: Crossover (Fig.~\ref{fig:flow}.3(c)), Top-$k$ selection, and Mutation (Fig.~\ref{fig:flow}.3(d)), produce the optimized circuit (Fig.~\ref{fig:flow}.4).  \emph{To our knowledge, this is the first application of evolutionary strategies to $T$-depth reduction.} The major contributions of this paper are defined as follows:

% In this paper, we develop a comprehensive framework for optimizing the quantum circuit by merging the layers of commutating $T$ gates to reduce the qubit overhead of magic state distillation and overall quantum error correction. We define the problem of merging layers of $T$ gates as a search optimization problem and design a Genetic Algorithm (GA) to approximate optimal solutions within the search space to effectively merge the layers and obtain a maximally optimized circuit. As observed in Fig. \ref{fig:flow}.3, we obtain the Clifford-pruned circuit as the initialization for the GA. The pairs of columns (layers of $T$ gates) are chosen based on their individual $T$-gate density to be merge pairs for the population of the GA as shown in Fig. \ref{fig:flow}.3(b). The GA operators, viz., crossover (Fig. \ref{fig:flow}.3(c)), a top-$k$ selection, and mutation (Fig. \ref{fig:flow}.3(d)) work on the merge pairs to obtain an optimally merged circuit (Fig. \ref{fig:flow}.4). A detailed description of the GA is presented in Section IV. \emph{To the best of our knowledge, this is the first attempt to optimize quantum circuits for reducing $T$-depth using evolutionary strategies.} The major contributions of this paper are defined as follows:
\begin{enumerate}
    \item We represent $T$-depth reduction as a search optimization problem, thereby providing an approximation for the optimally merged circuit.
    \item We propose expansion of $T$-gate layers in the quantum circuit to create better merge options, and a greedy algorithm for the selection of the initial layers for the merge, thus reducing the effective search space.
    \item We design a GA to explore optimal merge pairs in the reduced space to obtain better-merged circuits quickly.
    \item We evaluate the performance of our model on our benchmark with varying $T$-gate densities and obtain a $\sim21\%$ overall improvement in $T$-depth reduction over the state-of-the-art lookahead algorithm \cite{chatterjee2025artoptimizingtdepthquantum}. We also test our method on the reversible circuit benchmark \cite{tpar} \cite{cheung2009designoptimizationquantumpolynomialtime}.

\end{enumerate}

Section II provides background on FTQC and $T$-gate optimization. Section III introduces the problem statement. Section IV presents the proposed idea. We evaluate the idea in Section V and conclude the paper in Section VI.
\section{Background}

\subsection{Workflow of QEC design in FTQC}
In an FTQC stack, the user-designed quantum circuit must first be brought into a form suitable for error-corrected execution. We describe the flow in Fig. \ref{fig:flow}. \textbf{Logical synthesis} begins with the user’s high-level circuit description, which is Fig. \ref{fig:flow}(a), compiled into the native gate set of the target hardware, Fig. \ref{fig:flow}(b), and then further mapped into the universal \emph{Clifford+$T$} basis Fig. \ref{fig:flow}(c) required by surface-code QEC.
Next, all Clifford gates (single- and multi-qubit Pauli-product rotations) are systematically \textbf{commuted} through the remaining non-Cliffords by applying the Pauli-rotation commutation rules (if two Paulis commute they swap freely; if they anticommute, a phase factor is introduced \cite{littinski}), which pushes every \emph{Clifford} gate to the circuit’s end and gets absorbed into measurement frames.  What remains after this \emph{Clifford pruning} is a sequence of pure Pauli $\pi/8$ rotations, i.e., the $T$-gates (non-\emph{Clifford}), arranged in commuting layers.

This Clifford-free, columnar $T$-gate representation is the direct input to our $T$-depth optimization framework. In parallel, each logical $T$ is implemented fault-tolerantly via \emph{magic-state distillation}, producing high-fidelity $\lvert T\rangle$ ancillas, and injected into the circuit.  Finally, during execution the surface-code’s syndrome measurements are continuously decoded to identify and correct errors, completing the FTQC pipeline.

\subsection{What is $T$-depth Reduction?}
$T$-depth is the number of sequential layers of non-Clifford $\pi/8$ rotations ($T$ gates) in a circuit. Reducing $T$-depth packs commuting $T$ gates into fewer layers, directly reducing the required number of fresh high-fidelity \(\lvert T\rangle\) ancillas, since each layer consumes new magic states. However, the preparation of magic states is not trivial and involves hundreds of physical qubits and multiple rounds of error detection and correction~\cite{Bravyi_2005, haah2018codes, littinski}. Lower $T$-depth therefore reduces magic-state factory throughput, qubit overhead, and execution time, addressing the primary bottleneck of resource-intensive distillation in surface-code FTQC. Optimization of both $T$-count and $T$-depth can be addressed at two distinct stages in the FTQC stack. First, during logical synthesis the total number of Clifford and $T$ gates can be minimized via advanced circuit‐level decomposition and resource‐aware routing. Second, after Clifford pruning, when only non-Clifford Pauli $\pi/8$ rotations ($T$-gates) remain, several commutation and layering optimizations can be implemented to pack $T$ gates into as few layers as possible, thereby directly reducing the $T$-gates in the circuit and indirectly reducing the magic state consumption and overhead. 

Previous works have studied the idea of advanced circuit optimization in detail to present techniques to reduce the Clifford gates and $T$-gates. Authors in \cite{tpar} frame the $T$-depth minimization as finding the smallest number of independent sets in a matroid whose ground set is the circuit’s phase gadgets, leveraging Edmonds’ matroid-partitioning algorithm to optimally parallelize commuting phase rotations into minimal layers. This yields up to an average reduction of $\sim 40\%$ in the $T$-count and an average reduction of $\sim 61\%$ in the $T$-depth by zero ancilla usage, while simultaneously providing the theoretical upper bound for the optimisation in the $T$-depth in the presence of unlimited ancilla qubits. The technique of converting quantum circuits into ZX-calculus gadgets using the $T$-gate phases, and introducing a novel phase-teleportation procedure has also been studied in detail in \cite{pyzx}. They non-locally combine and cancel the $T$-gate gadgets without altering the circuit structure to obtain up to a $\sim 50 \%$ reduction in the $T$-count over the then State-of-the-art. TRASYN, a tensor-network-guided synthesis algorithm, was introduced in \cite{tensor}, which directly compiles arbitrary single-qubit unitaries in a Clifford+$T$ gate set, bypassing separate $R_z$ decompositions to yeild a $T$-count reduction of $3.5\times$ and a Clifford reduction of $\sim 7$ times. Recent work \cite{chatterjee2025artoptimizingtdepthquantum} addresses post-Clifford \(T\)-gate optimization via a limited-lookahead brute-force merging heuristic, which suits sparse but not \(T\)-dense circuits.  We instead cast \(T\)-depth reduction as a constrained optimization and employ Genetic Algorithms to efficiently approximate near-optimal merges in the resulting nonconvex search space.

\section{Problem Formulation}

\subsection{Problem Statement}

% In an FTQC workflow, the user-designed quantum circuits are transpiled to the native gate set of the quantum hardware and then converted to the Clifford+$T$ gate set before mapping them onto standard quantum error correction protocols like surface codes. The Clifford gates are absorbed by certain measurement-based computing procedures, and the $T$-gates require magic state distillation protocols to be pruned. After the absorption of the Clifford gates, the circuit gets transformed into a grid-like structure with $T$-gates dominating the columns across the qubits of the circuit. A singular magic state is consumed to prune one such column, which is an expensive procedure since, for near-term Fault-Tolerant Quantum Computers, since the number of columns can vary up to $\sim1000$ and the number of qubits for the magic state distillation increases the qubit overhead to $10^6$ \cite{littinski}. 
\begin{figure}
    
    \centering
    \includegraphics[width=\linewidth]{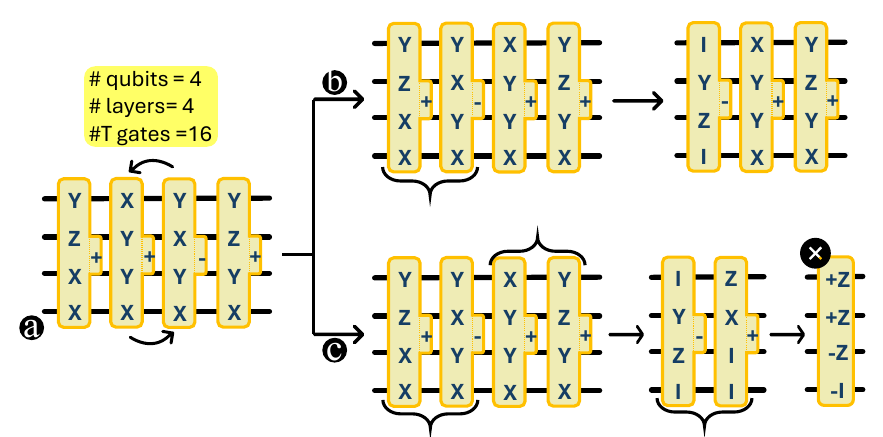}
    %\vspace{-10pt}
    \caption{A diagrammatic representation of the flow of $T$-depth reduction by merging layers of commutating $T$ gates. Part (a) represents the post-Clifford circuit with four $T$-gate layers. In part (b), after swapping columns 2 and 3, merging columns 1 \& 2 cuts $T$-depth by 25\% (magic states drop to three). However, in part (c), further merging columns 3 \& 4 is attempted, but fails at the final step due to a phase mismatch among the individual $T$-gates. }
    \label{fig:eg}
    %\vspace{-10pt}
\end{figure}

In Fig. \ref{fig:eg}, we demonstrate the idea of merging and reordering the $T$-gate layers based on commuting gates to efficiently combine them, thus reducing the depth while maintaining logical equivalence. 

\begin{lemma}[]\label{lem:mer} If the $T$-gate layers of a circuit commute, they can be merged and re-ordered within the circuit without loss of logical equivalence.
\end{lemma}
\begin{proof}
Let 
\(
U_j=e^{\,iA_j},
\quad
A_j = \sum_{q=1}^n \frac\pi8\,P_q^{(j)},
\quad j=1,\dots,m, \quad P_q\in\{X,Y,Z\}
\)
and assume \([A_j,A_k]=0\) for all \(j,k\). We define
\(
U_{\rm merged}
= \prod_{q=1}^n
  \bigl(R_{P_q^{(1)}}(\tfrac\pi8)\,\cdots\,R_{P_q^{(m)}}(\tfrac\pi8)\bigr)
= \exp\bigl[i(A_1+\cdots+A_m)\bigr].
\)
Therefore, we prove that the operations, Merging: 
\(
      U_{\rm merged}
      = U_m\,U_{m-1}\,\cdots\,U_1
    \), and Reordering (for any permutation \(\sigma\) of \(\{1,\dots,m\}\)) : 
    \(
      U_{\sigma(1)}\,U_{\sigma(2)}\,\cdots\,U_{\sigma(m)}
      = U_1\,U_2\,\cdots\,U_m
    \), preserve the logical equivalence of the quantum circuit.
    
We proceed by induction on \(m\ge2\).
\noindent\textit{Base case \((m=2)\).}
We have \(U_j=e^{\,iA_j}\) with \([A_1,A_2]=0\).  The Baker–Campbell–Hausdorff (BCH) series \cite{hall2015lie}
\(
  e^{iA_1}e^{iA_2}
  = e^{(iA_1 + iA_2 + \tfrac12[iA_1,iA_2] + \cdots)}
\)
collapses to
\(
  e^{iA_1}e^{iA_2}
  = e^{\,i(A_1+A_2)}
  = e^{iA_2}e^{iA_1}.
\)
Therefore, we obtain
\(
  U_{\rm merged}^{(2)}
  = \prod_{q=1}^n R_{P_q^{(1)}}(\tfrac\pi8)\,R_{P_q^{(2)}}(\tfrac\pi8)
  = e^{\,i(A_1+A_2)}
  = U_2\,U_1
  = U_1\,U_2.
\)

\noindent\textit{Inductive step.}
Let the claim hold for \(m\) columns:
\(
  U_{\rm merged}^{(m)}
  = U_m\,\cdots\,U_1
  = \exp\bigl[i(A_1+\cdots+A_m)\bigr],
\)
and is invariant under any permutation of the factors.  Adding the \((m+1)\)-th column \(U_{m+1}=e^{\,iA_{m+1}}\), since \([A_{m+1},A_1+\cdots+A_m]=0\), the BCH series \cite{hall2015lie} again yields
\(
  U_{m+1}\,U_{\rm merged}^{(m)}
  = e^{\,iA_{m+1}}\,e^{\,i(A_1+\cdots+A_m)}
  = e^{\,i(A_1+\cdots+A_{m+1})}
  = U_{\rm merged}^{(m+1)}.
\)
From above, we obtain 
\(
  U_{\rm merged}^{(m+1)} = U_{m+1}\,U_{\rm merged}^{(m)} = U_{m+1}\,U_m\,\cdots\,U_1.
\)
% \noindent\textit{(b) Reordering including the new column.}
Since, $A_{m+1}$ commutes with each $A_j$, we can swap $U_{m+1}$ past any subset of the existing $m$ factors (each swap justified by the base $m=2$ case), showing that for any permutation $\tilde\sigma$ on $\{1,\dots,m+1\}$,
\(
  U_{\tilde\sigma(1)}\,\cdots\,U_{\tilde\sigma(m+1)}
  = U_1\,U_2\,\cdots\,U_m\,U_{m+1}.
\)
\end{proof}
Building on Lemma \ref{lem:mer}, we now formalize our optimization objective. Our aim is to minimize the circuit’s $T$-depth by merging $T$-gate layers wherever the corresponding Pauli operators commute on each qubit. Any merge must preserve the global phase consistency of the resulting layer. We therefore cast the task as a combinatorial search: given the circuit’s inherent layer-wise arrangement of $T$-gates, the framework systematically explores admissible merge patterns and selects the one that yields the greatest reduction in $T$-depth while respecting the phase‐consistency constraint.

% Fig.~\ref{fig:eg} presents a 4-qubit, 4-column circuit comprising 16 $T$ gates and shows three representative strategies for layer formation. Given 4 columns, there exists $4! = 24$ possible permutations of column arrangements, from which we highlight three examples.
% In the first configuration (a), the circuit maintains its original ordering and structure, resulting in 4 distinct layers. The second configuration (b) rearranges the columns and successfully merges columns 1 and 2, thereby reducing the circuit depth to 3 layers. The third configuration (c) adopts the same column ordering as (b) but further compresses the circuit to 2 layers by additionally merging columns 3 and 4 alongside the earlier merge of columns 1 and 2.
% It further considers a hypothetical merge of the resulting columns; however, this operation fails due to incompatible $T$-gate phases within the combined column.
% This example underscores the value of examining alternative column permutations, as the mergeability of columns—and consequently the achievable circuit depth—is highly sensitive to their arrangement.

\subsection{Research Challenges}
By analyzing the problem of $T$-depth optimization as a search problem, we observe two main challenges: (1) The entire search space is \emph{exponentially} large. For a circuit with $c$ columns, the total number of possible merge pairs is a subset $M \subset$ $c\choose2$. The search will involve finding a set of disjoint pairs from $M$ which is of the order $O(2^{|M|})$ and practically unfeasible for quantum circuits with $\sim100$ qubits and $\sim1000$ columns; (2) Non-uniform $T$-gate densities make random or unguided merging ineffective: high-density columns resist merging and low-density columns yield negligible benefit, so blind pairing often increases overhead.

To address the exponential search space, we first apply a greedy filter, scoring each valid pair by combined $T$-count and density gap, and then run an iterative GA on this pruned set.  To cope with heterogeneous $T$-gate density, our expansion policy and GA seeding deliberately match high‐density with low‐density columns, so merges satisfy $T_{max}$ and phase constraints while minimizing padding overhead.

\section{Optimizing T gates}

\subsection{Expanding for Efficient Merging}
To introduce a global reordering perspective for our Genetic Algorithm, we propose to split the $T$-gate layers to provide a finer granularity for the merging procedure.

\begin{proposition}[]\label{prop:1}
If we partition $T$-layers in a quantum circuit, the resulting circuit is still logically consistent.
\end{proposition}
\begin{proof}
Building on the same setup from Lemma \ref{lem:mer}, 
we partition $\{1,\dots,n\}$ into $k$ disjoint subsets $S_1,\dots,S_k$, and define
\(
  U_j = \prod_{q\in S_j} R_{P_q}\bigl(\tfrac\pi8\bigr)
      = e^{\,iA_j},
  \quad
  A_j = \sum_{q\in S_j}\frac\pi8\,P_q.
\)
Then for any $k\ge1$ and any permutation $\sigma$ of $\{1,\dots,k\}$, we prove, 
\(
  U = U_1\,U_2\cdots U_k
    = U_{\sigma(1)}\,U_{\sigma(2)}\cdots U_{\sigma(k)}.
\)
Since the subsets $S_j$ are disjoint, each
\(
  A_j = \sum_{q\in S_j}\frac\pi8\,P_q
\)
acts on a distinct set of qubits, so
\(
  [A_j,A_{j'}] = 0
  \quad\forall\,j\neq j'.
\)
Hence for any two indices $j,j'$ the BCH series \cite{hall2015lie} collapses to 
\(
  e^{iA_j}e^{iA_{j'}}
  = e^{\,i(A_j + A_{j'})}
  = e^{iA_{j'}}e^{iA_j}.
\)
By iterating this merge across all $k$ factors, we get
\(
  \prod_{j=1}^k U_j
  = \prod_{j=1}^k e^{\,iA_j}
  = e^{(i\sum_{j=1}^k A_j)}
  = e^{iA}
  = U.
\)
Moreover, since each $U_j$ commutes with every other $U_{j'}$, any permutation of the factors leaves the product unchanged:
\(
  U_{\sigma(1)}\,U_{\sigma(2)}\cdots U_{\sigma(k)}
  = \prod_{j=1}^k U_j
  = U.
\)
\end{proof}
\begin{algorithm}[h]
\caption{Greedy Merge Filter}
\label{algo:greedy}
\begin{algorithmic}[1]
\Require Circuit $C=[c_1,\dots,c_n]$, candidate pairs $P$, thresholds $T_{\max},\delta_{\max},k_{\min}$, weight $\beta$
\Ensure Selected merges $S$
\If{$|P|\le k_{\min}$} \Return $P$ \EndIf
\State For each column $i$: $T_i\leftarrow\#\{T\text{-gates in }c_i\}$ and $D_i\leftarrow T_i/|c_i|$
\State $\mathsf{ScoredPairs}\leftarrow\bigl\{(i,j,\;1-|D_i-D_j|+\beta\,(T_{\max}-T_i-T_j))\mid(i,j)\in P,\;|D_i-D_j|\le\delta_{\max},\;T_i+T_j\le T_{\max}\bigr\}$
\State Sort $\mathsf{ScoredPairs}$ by descending score
\State $S\leftarrow\emptyset,\;U\leftarrow\emptyset$
\For{each $(i,j,\_)$ in $\mathsf{ScoredPairs}$}
  \If{$\{i,j\}\cap U=\emptyset$}
    \State add $(i,j)$ to $S$ and $\{i,j\}$ to $U$
  \EndIf
\EndFor
\State\Return $S$
\end{algorithmic}
\end{algorithm}
Splitting a $T$-gate column along qubit lines and padding with identities while preserving its overall phase can enhance mergeability in circuits of moderate or high $T$-density. From Proposition \ref{prop:1}, we infer that splitting the $T$-gate layers does not mess with the logical sense of the circuit. Prior work \cite{chatterjee2025artoptimizingtdepthquantum} confirms this empirically but applies ad hoc expansion, often bloating circuit size.  We instead prescribe an expansion factor driven by the local $T$-density
\(
\gamma = \frac{a+b+c}{n},
\)
where $n$ is the qubit count and $a,b,c$ are the numbers of $R_x,R_y,R_z$ gates in the column.  Under this rule, very sparse columns undergo minimal expansion (avoiding identity-heavy segments), and already dense columns receive little or no expansion (avoiding needless overhead).
To balance these trade-offs, we define a column-dependent expansion factor $E(C)$, scaled by the density ratio $\gamma$, such that merging remains computationally tractable:
\(
E(C) = \log(n+1)\,(1 - \gamma^2)\,\gamma^{1+\gamma} + \left\lceil\frac{a + b + c}{\tau}\right\rceil,
\)
where the scaling term $\tau$ is defined as
\(
\tau = \max\left(1,\, n^{\alpha}e^{2\gamma}\right),
\)
and $\alpha$ is a tunable hyperparameter. The logarithmic term $\log(n+1)$ ensures that the expansion factor does not grow exponentially with circuit size and is lower-bounded by 1.
To ensure that the expanded columns are well-structured for merging, we further propose splitting them around qubits with higher local $T$-gate concentration. For a given column $C$ and qubit $i$, we define the local density measure $P_{i,C}$ as:
\(
P_{i,C}=g_{i,C}+\mu\Bigl(g_{i,C}-\tfrac{1}{\lvert N(i)\rvert}\sum_{j\in N(i)}g_{j,C}\Bigr)+(1-\gamma)\sum_{j\in N(i)}g_{j,C}.
\)
where $g_{i,C} \in \{0, 1\}$ indicates $1$ for a $T$-gate, and $0$ otherwise, at qubit $i$ in column $C$, and $\mu$ denotes the variance of gate placements across the column, capturing the clustering of $T$-gates. The neighborhood $N(i)$ is defined in
\(
\{ j \mid |j - i| \leq k,\ 0 \leq j < n \},
\)
where $k$ is a locality parameter determining the extent of qubit adjacency considered. Here, we keep the value of $k$ low ($<10$) to ensure a strict locality reference. To rank qubits for splitting, we define a splitting probability score:
\(
S_{i,C} = \frac{P_{i,C}}{\sum_j P_{j,C}},
\)
and select qubits in ascending order of $S_{i,C}$, prioritizing those with lower local $T$-gate density. This ensures that splits occur in regions with lower concentrations of non-identity operations, thus improving the likelihood of effective merges in subsequent optimization steps.

\begin{algorithm}[H]
\caption{Genetic Algorithm}
\label{algo:combinedGA}
\begin{algorithmic}[1]
\Require Circuit $C=[c_1,\dots,c_n]$, GA params $(N,G,k)$
\Ensure Reduced circuit $C'$, merges $M$
\State $C_{\rm curr}\!\gets\!C$, $M\!\gets\!\emptyset$
\Repeat
  \State $P\!\gets\!\{(i,j)\mid\mathsf{CanMerge}(c_i,c_j)\}$ in $C_{\rm curr}$
  \If{$P=\emptyset$} \textbf{break} \EndIf
  \State int. population of $N$ non‐overlapping subsets of $P$
  \State $B\!\gets\!\emptyset$
  \For{$1\le t\le G$}
    \State Score and select top $k$; generate new population via crossover + mutation
    \State Update $B$ if best improves
  \EndFor
  \If{$B=\emptyset$} \textbf{break} \EndIf
  \State $U\!\gets\!\bigcup_{(i,j)\in B}\{i,j\}$; \;$C_{\rm next}\!\gets\![]$
  \For{$(i,j)\in B$} \State append Pauli‐product of $c_i,c_j$ to $C_{\rm next}$ \EndFor
  \For{$k\notin U$} \State append $c_k$ to $C_{\rm next}$ \EndFor
  \State $C_{\rm curr}\!\gets\!C_{\rm next}$; \;$M\!\gets\!M\cup B$
\Until{no new merges}
\State \Return $(C_{\rm curr},M)$
\end{algorithmic}
\end{algorithm}

\subsection{Generating Suitable Merge Patterns}
GAs typically start from a random population of merge‐pair candidates and evolve it to maximize valid merges.  However, random initialization can trap the GA in poor local minima.  To avoid this, we employ a structured pre‐selection based on each column’s $T$‐gate density $\gamma$.  Since dense columns are harder to merge, we sort all columns by $\gamma$ and form initial candidates by pairing those with the highest density with those with the lowest.  This targeted seeding improves both the diversity and quality of the starting population of the GA.
To reduce the search space, we compute for each valid pair \((i,j)\) its total \(T\)-count \(T_{\rm sum}\) and density gap \(\Delta_D\), then assign the score
\(
\mathrm{score}(i,j)=1-\Delta_D+\beta\,(T_{\max}-T_{\rm sum}),
\)
with \(\beta\in[0,1)\) favoring lower \(T_{\rm sum}\).  Algorithm~\ref{algo:greedy} then greedily selects non‐overlapping pairs in descending score order.  With \(O(n^2)\) candidates, the initial \(\gamma\)-based sort takes \(O(n^2\log n)\), and the subsequent matching over \(O(n^2)\) pairs costs \(O(n^2\log n^2)\).

% \begin{algorithm}[H]
% \caption{Iterative Genetic Algorithm for Column Reduction}
% \label{algo:GA1}
% \begin{algorithmic}[1]
% \Require Circuit $C = [c_1, c_2, \dots, c_n]$, Initial merge candidates $P_0$
% \Ensure Reduced circuit $C'$, final merge plan $M$

% \State $C_{\text{curr}} \gets C$, $M \gets \emptyset$, $r \gets 0$

% \While{true}
%     \State $P_r \gets$ all valid merge pairs $(i,j)$ in $C_{\text{curr}}$ where \Call{CanMerge}{$c_i$, $c_j$}
%     \If{$P_r = \emptyset$} \textbf{break} \EndIf

%     \State $B_r \gets$ \Call{RunGeneticAlgorithm}{$C_{\text{curr}}$, $P_r$}
%     \If{$B_r = \emptyset$} \textbf{break} \EndIf

%     \State $U \gets$ set of indices used in $B_r$
%     \State $C_{\text{next}} \gets []$
%     \For{each $(i,j) \in B_r$}
%         \State $m \gets$ column-wise Pauli product of $c_i$, $c_j$
%         \State Append $m$ to $C_{\text{next}}$
%     \EndFor
%     \For{each $k$ not in $U$}
%         \State Append $c_k$ to $C_{\text{next}}$
%     \EndFor

%     \State $C_{\text{curr}} \gets C_{\text{next}}$
%     \State $M \gets M \cup B_r$, $r \gets r + 1$
% \EndWhile

% \State \Return $(C_{\text{curr}}, M)$
% \end{algorithmic}
% \end{algorithm}

\subsection{Search for the Best Possible Merges}

To fully optimize the $T$-depth of a circuit, the proposed GA iteratively searches for the best set of non-overlapping merge pairs until no further beneficial merges can be found.
\emph{Initialization: }
A population of size $N$ is initialized from the filtered pool of merge candidates, ensuring that all selected pairs in a chromosome are non-overlapping. Let $\mathcal{P} = \{\mathcal{M}_1, \mathcal{M}_2, \dots, \mathcal{M}_N\}$ denote the population of $N$ chromosomes in the current generation.
\emph{Evaluation: }
Each chromosome is evaluated using a fitness function that quantifies the number of valid merges it encodes. Formally, for a circuit $C = [c_0, c_1, \dots, c_n]$ and a chromosome $\mathcal{M} = \{(i_1, j_1), (i_2, j_2), \dots\}$, the fitness function iterates through each candidate pair $(i, j)$ and checks two conditions: (i) neither $i$ nor $j$ has been used in a prior merge in $\mathcal{M}$ (ensuring the non-overlapping constraint), and (ii) the columns $c_i$ and $c_j$ are mergeable according to the Pauli product rules, i.e., all element-wise products result in a column with uniform global phase.
\(
\text{Fitness}(\mathcal{M}, C) = \sum_{(i,j) \in \mathcal{M}} \mathbf{1}_{\text{valid}(i,j)}
\)
where the indicator function $\mathbf{1}_{\text{valid}(i,j)}$ is defined as:
\(\mathbf{1}_{\mathrm{valid}(i,j)}=1\) exactly when \(i,j\notin U\) and \(\mathsf{CanMerge}(c_i,c_j)\) is true, and \(0\) otherwise, and, $U$ represents the set of column indices that have already been used in previous merges within the chromosome:
\(
U = \bigcup_{\substack{(i', j') \in \mathcal{M} \\ (i', j') \text{ processed before }(i,j)}} \{i', j'\}.
\)
\emph{Selection: }
Following evaluation, the algorithm enters the \textit{selection phase}, which balances \emph{elitism} (retaining the best-performing chromosomes) with \emph{diversity} (exploring new areas of the solution space).
\textbf{Ranking:} All chromosomes in $\mathcal{P}$ are sorted in descending order of $\text{Fitness}(\mathcal{M}_i, C)$.
\textbf{Top-$k$ Retention:} The top $k$ chromosomes, denoted $\mathcal{P}_{\text{elite}} = \{\mathcal{M}_1^*, \dots, \mathcal{M}_k^*\}$, are retained without modification and used as \emph{parents} for generating the next generation.
\textbf{Offspring Generation:} The remaining $N - k$ individuals in the next generation are produced by applying crossover and mutation to randomly selected pairs from $\mathcal{P}_{\text{elite}}$.
This top-$k$ elitism strategy ensures that the best chromosome discovered across all generations is preserved and returned as the final merge plan for the current iteration.
\emph{Crossover and Mutation: }
After selecting the top chromosomes, we perform \emph{crossover} by combining one half of the parent \(\mathcal{M}_a^*\) with the complementary half of \(\mathcal{M}_b^*\), then removing any overlapping pairs to enforce non‐overlap. To preserve diversity, we employ a reset‐style \emph{mutation}: with a small probability, the offspring is discarded and replaced by a new, randomly generated valid chromosome from the candidate pool.  This reset mutation injects structural variation, mitigating premature convergence while maintaining overall search convergence.

\section{Results}

\subsection{Simulation Setup}
All the algorithms for this work have been implemented in Python 3.12.1 on an Intel Core i9-13900K CPU with a clock frequency of 4.8GHz. 
\begin{figure}
    \centering
    \includegraphics[width=0.8\linewidth]{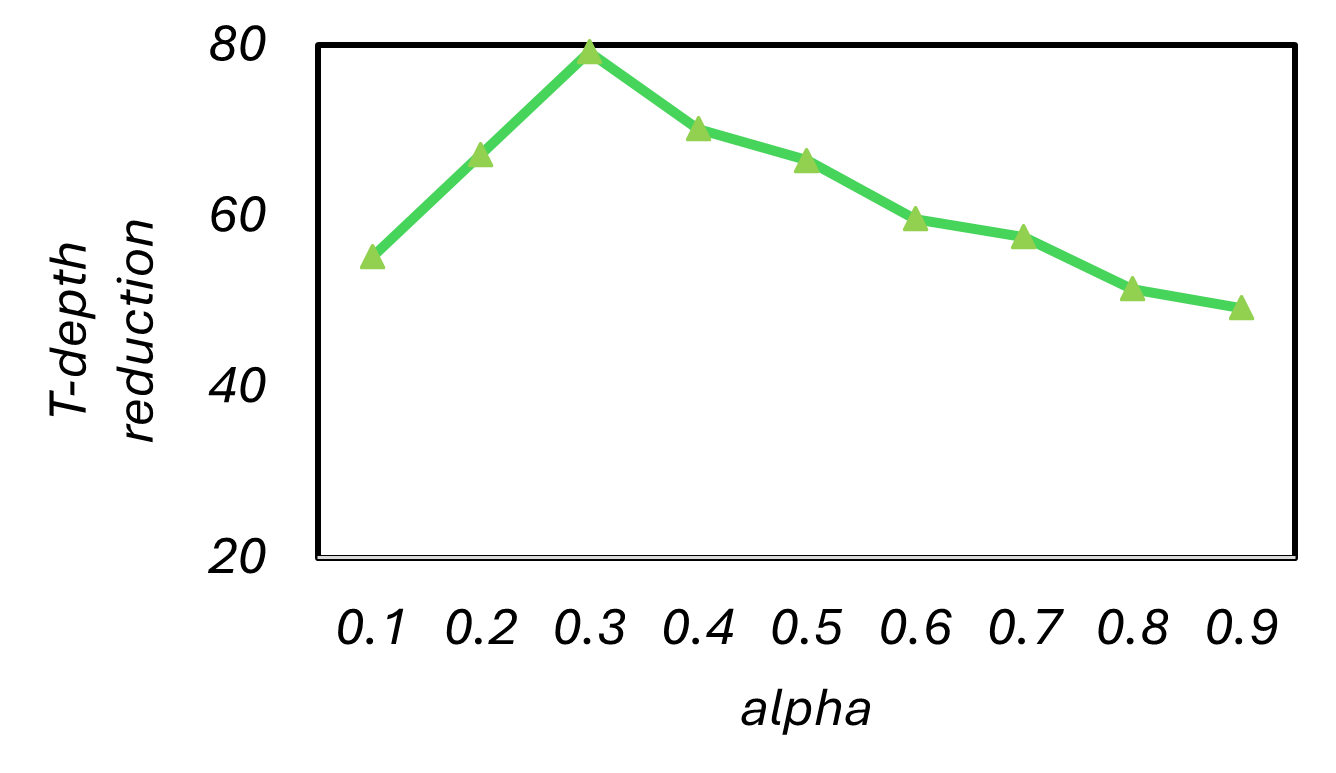}
    \vspace{-10pt}
    \caption{In this plot we determine the overall $T$-depth reduction across different values of $\alpha$. These experiments have been done on large circuits ($\sim 90-100$ qubits) with a high $T$-gate density. We can clearly see that $\alpha=0.3$ gives the best results for the expansion overhead.}
    \label{fig:alpha}
    \vspace{-10pt}
\end{figure}

\begin{figure*}
    \centering
    \includegraphics[width=0.9\linewidth]{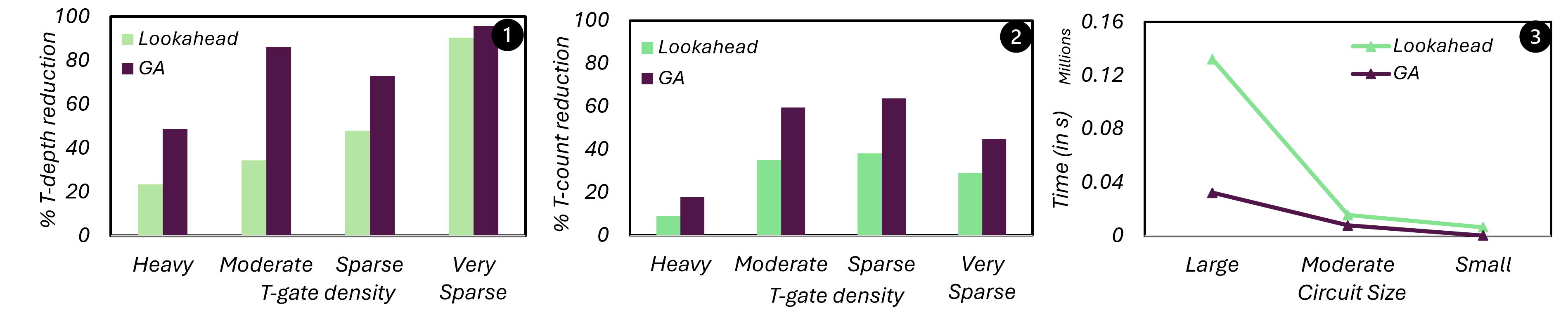}
    \vspace{-10pt}
    \caption{In (1) we observe the comparison in the reduction in $T$-depth between our proposed GA and the lookahead approach. We observe over 100\% increase in the reduction for circuits with heavy $T$-gate density, and an almost 3$\times$ improvement for moderately dense circuits. In (2) we observe the comparison in the reduction in $T$-count between our proposed GA and the lookahead approach.On comparing, we observe an average improvement of $\sim1.75\times$ for the circuits with the maximum impact being observed in the heavily dense circuits ($\sim2.35\times$). Both represent results taken on large circuits ($\sim 90-100$ qubits). In (3) we observe the runtime for $1000$ circuits profiled across three circuit sizes (small (10–20 qubits), moderate (60–70 qubits), and large (90–100 qubits)) is shown in this plot. GA has a $\sim4\times$ implementation advantage for large circuits and is significantly better for the other two profiles as well.}
    \label{fig:allplot}
    %\vspace{-10pt}
\end{figure*}
\textbf{Hyperparameters: }For the expansion factor described in Section IV.A, the value of $\alpha$ is set to 0.3 for our experiments. We empirically find that $\alpha=0.3$ is the best for expanding the columns to increase the mergeability of the columns without creating a significant overhead in the problem size (Fig. \ref{fig:alpha}). 
For the greedy selection of initial merge pairs, we choose a relatively high weight factor, $\beta$ equals 0.8. We run the GA for 20 generations with an initial population size of 20 and a low mutation rate of 0.2. For the top-$k$ selection, we set the value of $k$ to be 4 (top 20\% of the population). The population size can be increased to accommodate a larger number of merge pairs, however, we observe the number of generations to be enough for optimal convergence.
\textbf{Benchmarking: }
We first evaluated our pipeline on the standard Clifford + $T$ benchmarks from \cite{tpar}\cite{pyzx}—VBE and QCLA adders, CSLA-MUX, and GF\((2^m)\) multipliers (5–96 qubits, up to $\sim 2400$ \(T\)-gates)—but, given their Toffoli-dominated structure, we also generated synthetic circuits as discussed in \cite{chatterjee2025artoptimizingtdepthquantum}.  Each synthetic instance is specified by qubit count \(n\), column count \(c\), and total \(T\)-gates ($\leq (n,c)$), with \(T\)-gates placed uniformly at random across columns (capped at one per qubit) and axes drawn from \(\{X,Y,Z\}\), while each column’s global phase is chosen via Bernoulli \((p=0.5)\).

\subsection{Performance Evaluation}
We consider three primary resource metrics while evaluating against the current SOTA: (i) Reduction in $T$ gate count; (ii) Reduction in $T$-depth; and (iii) Overall runtime of the algorithms. These metrics give an overall idea of the range of optimization achieved in circuits and the reduction in the qubit overhead for standard QEC procedures like surface codes.
\textbf{$T$-depth: }
We observe the $T$-depth reduction in Table \ref{tab:comp}, where $T_d$ is the original value of $T$-depth, \cite{tpar}$_d$ is the $T$-depth obtained by the previous SOTA, and $T_d^{p}$ is the $T$-depth obtained by the proposed GA approach. To have an apples-to-apples comparison, we put our method up against the 0-ancilla implementation of \cite{tpar} as we propose the merging of $T$-gate layers without the use of extra ancillas. We find that our approach provides an average improvement of $\sim 2.58\times$ than \cite{tpar} without the use of extra qubits, which is key to the implementation in an FTQC setting. In comparison, the method in \cite{tpar} uses an option of ``infinite'' ancilla qubits to achieve $T$-depth reduction close to our proposed method, but it is a theoretical bound and cannot be achieved due to engineering constraints. 
On comparing with the lookahead algorithm \cite{chatterjee2025artoptimizingtdepthquantum} (23.57\%), we find a reduction of 48.72\% in $T$-depth for large ($\sim$90-100 qubits) heavily $T$-gate dense circuits ($\sim100\%$ $T$-gates). Also for mid-sized circuits ($\sim$50-60 qubits), we achieve a $\sim 22\%$ improvement. Overall, we obtain a 79.2\% $T$-depth reduction on large circuits. A graphical evaluation with the lookahead method over varying $T$-gate densities in large circuits is presented in Fig. \ref{fig:allplot} (1). 
\textbf{$T$-count: }
In addition to reducing the $T$-depth, we evaluate our method’s effectiveness in minimizing the overall $T$-count—i.e., the total number of $T$-gates remaining after optimization. We improve by $\sim 4.5 \times$ over \cite{tpar} and $\sim 1.55 \times$ over \cite{pyzx} on average. The column labeled $T_d$ shows the original $T$-depth, and $T_d^{p}$ shows the optimized depth using our method in Table \ref{tab:comp}. These results demonstrate the dominance of our approach over existing synthesis methods due to the use of zero ancilla qubits. Compared to the lookahead method, we achieve 18.01\% reduction in dense large circuits, versus only 8.97\%. Overall, we observe a 41.9\% $T$-depth reduction on large circuits. A graphical comparison over varying $T$-gate densities appears in Fig. \ref{fig:allplot} (2).
\textbf{Runtime analysis: }
A single run of our genetic algorithm (GA) over one generation requires evaluating and evolving a population of size \(N\) across all \(n\) columns and \(m\) candidate pairs, for a cost of
\(
\mathcal{O}\bigl(N\cdot (n + m)\bigr)\quad\text{per generation,}
\)
and hence
\(
\mathcal{O}\bigl(G\cdot N\cdot (n + m)\bigr)
\)
for \(G\) generations.  In our iterative pipeline, both \(n\) and \(m = O(n^2)\) decrease each round, so although the worst-case bound for \(R\) rounds is
\(
\mathcal{O}\bigl(R\cdot G\cdot N\cdot (n + m)\bigr),
\)
The actual wall-clock time remains in the order of hours on thousand-column circuits due to progressive circuit compression and greedy pre-filtering.
By contrast, the lookahead method of \cite{chatterjee2025artoptimizingtdepthquantum} partitions the \(n\) columns into windows of size \(k\) and exhaustively explores all \(k!\) reorderings per window.  Its runtime scales as
\(
\mathcal{O}\bigl(n\cdot k\cdot k! \cdot q \cdot \log n\bigr)
\)
for a \(q\)-qubit, \(n\)-column circuit.  To achieve strong global reductions they set \(k=14\), but since \(14!\approx8.7\times10^{10}\), execution can take days.  In contrast, our GA—with its score-based filtering and shrinking search space—completes in hours (Fig.~\ref{fig:allplot}(3)).

%\subsection{Parallelization in GA}
\begin{table}[ht]
\centering
\begin{tabular}{l|rrrrrrr}

Circuit (n) & $T_c$ & $T_d$ & \cite{pyzx}$_c$ & \cite{tpar}$_c$ & \cite{tpar}$_d$ & $T_{c}^{p}$ & $T_d^{p}$ \\
\hline \hline
adder$_8$ (24)           & 399  & 69  & 167 & 215  & 30  & 86  & 12  \\
rc-adder$_6$ (14)       & 77   & 33  & 47  & 63   & 22  & 45  & 11  \\
vbe-adder$_3$ (10)       & 70   & 24  & 24  & 24   & 9   & 16  & 4   \\
csla-mux$_3$ (15)        & 70   & 21  & 45  & 62   & 8   & 33  & 3   \\
csum-mux$_9$ (30)        & 196  & 18  & 72  & 112  & 9   & 58  & 3   \\
GF($2^4$)-Mult (12)  & 112  & 36  & 52   & 68   & 4   & 24  & 2   \\
GF($2^5$)-Mult (15)   & 175  & 48  & 86   & 111  & 5   & 30  & 2   \\
GF($2^6$)-Mult (18)   & 252  & 60  & 122   & 150  & 5   & 36  & 2   \\
GF($2^7$)-Mult (21)   & 343  & 72  & 173   & 217  & 7   & 42  & 2   \\
GF($2^8$)-Mult (24)   & 448  & 84  & 214   & 264  & 7   & 48  & 2   \\
GF($2^{16}$)-Mult (48) & 1856 & 180 & --  & 1040 & 12  & 96  & 2   \\
GF($2^{32}$)-Mult (96) & 7291 & 372 & --  & 4128 & 23  & 192 & 2   \\
ham$_{15}$-low (17)        & 161  & 44  & 97  & 97   & 19  & 97  & 19  \\
ham$_{15}$-med (17)        & 574  & 154 & 212 & 230  & 62  & 230 & 62  \\
ham$_{15}$-high (20)       & 2457 & 660 & 1013& 1019 & 274 & 1019& 274 \\
mod-mult$_{55}$ (9)     & 49   & 15  & 20  & 37   & 7   & 19  & 4   \\
mod-red$_{21}$ (11)       & 119  & 48  & 73  & 73   & 25  & 58  & 17  \\
mod $5_4$ (5)        & 28   & 12  & 7   & 16   & 6   & 7   & 3   \\
qcla-adder$_{10}$ (36)    & 589  & 24  & 158 & 162  & 11  & 158 & 6   \\
qcla-com$_{7}$ (24)        & 203  & 27  & 91  & 95   & 12  & 87  & 7   \\
qcla-mod$_{7}$ (26)        & 413  & 57  & 216 & 249  & 29  & 186 & 14  \\
\hline
\end{tabular}
\caption{Comparison of our proposed method over the reversible circuit benchmarks}
\label{tab:comp}
\vspace{-20pt}
\end{table}

\subsection{Viability in FTQC}
In standard compilation, hardware‐specific Clifford optimizations, e.g., patch compression for surface codes or pruning on IBM’s heavy-hex and QLDPC architectures, are applied before tackling non-Clifford gates.
Our GA operates on Pauli \(\pi/8\) layers in the post‐Clifford, pre‐distillation stage and makes no assumptions about the underlying code or connectivity. By reducing both \(T\)-depth and \(T\)-count, it directly reduces magic‐state demand and eases scheduling in any magic‐state factory.
Consider IBM’s \textit{ibm\_brisbane} (127 qubits, readout error \(\approx1.709\times10^{-2}\)).  To keep logical‐qubit failure \(<1\%\) and per‐\(T\)-gate error \(<1\%\), the 116\(\to\)12 distillation protocol (error \(\approx41.25\,p^4\)) requires  
\(
41.25\,p^4 \le10^{-2}
\quad\Longrightarrow\quad
p\lesssim3.52\times10^{-6}.
\)
For circuits with \(T\)-depth \(\sim10^3\) and \(T\)-count \(\sim10^5\), compounded \(T\)-gate errors force \(p\sim10^{-7}\), which is impractical.  Our method can reduce \(T\)-depth to \(\sim10^2\) and \(T\)-count to \(\sim10^4\), loosening the requirement to \(p\sim10^{-6}\)—within current hardware capabilities—thus making large‐scale FTQC more feasible.

\section{Conclusion}

We propose a hardware‐agnostic Genetic Algorithm for $T$-gate optimization at the post-Clifford, pre-distillation stage.  Modeling $T$-depth and $T$-count minimization as a constrained search, our GA merges commuting Pauli $\pi/8$ layers under phase and mergeability constraints, using a greedy $T$-density seeding for faster convergence.  On 1000 synthetic circuits (90–100 qubits), we achieve up to 79.2\% $T$-depth and 41.9\% $T$-count reductions—outperforming lookahead algorithms by up to $1.8\times$ in $T$-count, $1.2\times$ in $T$-depth, and delivering a $4\times$ speedup and a $2.6\times$ improvement in $T$-depth reduction on reversible‐circuit benchmarks.  Compatible with surface, heavy-hex, QLDPC, and trapped‐ion architectures, it substantially lowers magic-state overhead in fault‐tolerant pipelines.

\section*{Acknowledgment}

The work is supported in parts by the National Science Foundation (NSF) (CNS-1722557, CCF-1718474) and gifts from Intel.

\bibliographystyle{unsrt}
\bibliography{refs}

\end{document}